\documentclass[A4]{article}
\usepackage[T1]{fontenc}
\usepackage[latin1]{inputenc}
\usepackage{amsmath, amssymb, amsthm}
\usepackage[english]{babel}
\usepackage{indentfirst}
\usepackage{booktabs}
\usepackage{array}
\usepackage{caption,appendix}
\usepackage{geometry}
\usepackage{float}
\usepackage{cancel}
\usepackage{bbold}
\usepackage{authblk}
\usepackage{url}
\numberwithin{equation}{section}
\begin{document}
\title{\textbf{Weak field limit and gravitational waves in  higher-order
gravity}}
\author[1,2,3,4]{Salvatore Capozziello}
\author[5]{Maurizio Capriolo} 
\author[5]{Loredana Caso}
\affil[1]{\emph{Dipartimento di Fisica "E. Pancini", Universit\`a		   di Napoli {}``Federico II'', Compl. Univ. di
		   Monte S. Angelo, Edificio G, Via Cinthia, I-80126, Napoli, Italy, }}
		  \affil[2]{\emph{INFN Sezione  di Napoli, Compl. Univ. di
		   Monte S. Angelo, Edificio G, Via Cinthia, I-80126,  Napoli, Italy,}}
 \affil[4]{\emph{Laboratory for Theoretical Cosmology,
Tomsk State University of Control Systems and Radioelectronics (TUSUR), 634050 Tomsk, Russia,	}}
 \affil[4]{\emph{ Tomsk State Pedagogical University, ul. Kievskaya, 60, 634061
Tomsk, Russia, }}
\affil[5]{\emph{Dipartimento di Matematica Universit\`a di Salerno, via Giovanni Paolo II, 132, Fisciano, SA I-84084, Italy.} }
\date{\today} 

\maketitle
\noindent {\it We wish to dedicate this paper to the beloved memory of Maria Transirico who passed away while this paper was being completed. We mourn the loss of a mentor and lament the loss of a dear friend. Thanks Mariella!}
 
\begin{abstract}
We derive the  weak field limit for a gravitational Lagrangian density $L_{g}=(R+a_{0}R^{2}+\sum_{k=1}^{p} a_{k}R\Box^{k}R)\sqrt{-g}$ where higher-order derivative terms in the Ricci scalar $R$ are taken into account. The interest for this kind of effective theories comes out from the consideration of the infrared and ultraviolet behaviors of gravitational field and, in general, from the formulation of quantum field theory in curved spacetimes. Here, we obtain    solutions in weak field regime  both in vacuum and in the presence of matter and derive  gravitational waves considering the contribution of  $R\Box^{k}R$ terms. By using a suitable set of coefficients $a_{k}$, it  is possible to find up to $(p+2)$ normal modes of oscillation with six polarization states with helicity 0 or 2. Here $p$ is the higher order term in the $\Box$ operator appearing in the gravitational Lagrangian. More specifically: the mode $\omega_{1}$,  with $k^{2}=0$, has   transverse polarizations $\epsilon_{\mu\nu}^{\left(+\right)}$ and $\epsilon_{\mu\nu}^{\left(\times\right)}$ with helicity 2;  the $(p+1)$ modes $\omega_{m}$,  with $k^{2}\neq0$, have  transverse polarizations $\epsilon_{\mu\nu}^{\left(1\right)}$ and non-transverse ones $\epsilon_{\mu\nu}^{\left(\text{TT}\right)}$, $\epsilon_{\mu\nu}^{\left(\text{TS}\right)}$, $\epsilon_{\mu\nu}^{\left(L\right)}$ with helicity 0.
\end{abstract}
\noindent Keywords: Extended theories of gravity; weak field limit; gravitational waves. \\
\noindent Mathematics Subject Classification 2010: 83C40, 83C05, 83C10
\thispagestyle{empty}
\section{Introduction}
The need of extending  General Relativity arises from astrophysical and cosmological reasons like the impossibility of explaining the early and late time accelerated expansion of the Universe and  from the necessity to unify  gravitation with the others fundamentals interactions under the standard of quantum field theory. Indeed,  quantum effects intervene at  ultraviolet regimes and Standard Cosmological Model, based on  General Relativity,   does not work. On the other hand,  also at infrared scales, the Einstein theory  is not explaining  galactic, extra-galactic and cosmological structures, except by introducing  exotic ingredients like dark matter and energy, not probed, until now, at fundamental scales.  
It is worth pointing  out that dark energy and dark matter may be also unified under the same barotropic fluid, without necessarily invoking extensions of General Relativity as discussed in literature \cite{luongo,agostino,dunsby,aviles,quevedo,bravetti}.  
Approaches toward quantum cosmology as a source for dark energy have been proposed in view of observations in  some model-independent pictures \cite{mancini, luongo2,luongo3,luongo4,luongo5, dunsby2}. 

Besides the above approaches, Extended Theories of Gravity (ETGs) \cite{Ext Theory, CFA,sergey,vasilis} act on the geometry  considering higher order curvature invariants, like $R^{\mu\nu}R_{\mu\nu}$, $R^{\mu\nu\lambda\sigma}R_{\mu\nu\lambda\sigma}$, or higher order derivative terms into the Hilbert-Einstein  Lagrangian, linear in the Ricci scalar\footnote{Such invariants can be constructed starting from the Ricci tensor $R_{\mu\nu}$,  the Riemann tensor $R_{\mu\nu\lambda\sigma}$, the Weyl tensor $W_{\mu\nu\lambda\sigma}$, or the Gauss-Bonnet topological invariant \cite{felix} which is ${\cal G}=R^2-4R_{\mu\nu}R^{\mu\nu}+R_{\mu\nu\lambda\sigma}R^{\mu\nu\lambda\sigma}$. In general,  any Kretschmann curvature invariant can give rise to ETGs \cite{bini}.} $R$.  In particular, invariant  terms like $R\Box^{k}R$, where $\Box$ is the D'Alembert operator,  give rise to field equations    with derivative  orders higher than 2, namely of $(2k+4)$ order\footnote{It is worth saying  that $f(R)$ gravity, if $f(R)\neq R$,  is a fourth-order theory in metric representation.}. Specifically, this kind of contributions emerge in quantum field theory formulated on curved spacetimes \cite{birrell} and in effective actions adopted for  quantum gravity \cite{vilkovisky}.  Physical phenomena related to these terms,  specifically the light bending,  the weak-field limit and  ghost-free behavior are discussed in literature \cite{breno1,breno2,breno3}.

Considering possible effects at infrared scales, the idea is that further degrees of freedom coming from  geometrical invariants allow to avoid the  introduction of ad hoc fields (dark matter, dark energy, quintessence, etc.), explaining in a natural way  cosmological dynamics and astrophysical structures  like the inflation, the today observed acceleration of universe,   the flat rotation curves of galaxies, the structure of galaxy clusters \cite{annalen}.
However, the possibility to really test theories of gravity is recently emerged with the discovery of gravitational waves.
They, early predicted by Einstein's General Relativity,  were indirectly observed in 1982 thanks to the decrease of the orbital period observed for the binary pulsar PRS B1913+16, discovered in 1974 by Hulse and Taylor \cite{HT,TW}. 
On September 14th, 2015 at 09:50:45 UTC, the LIGO Hanford, WA, and Livingstone, LA, observatories detected the  gravitational-wave transient generated by  the merge of  a black hole binary system  \cite{ABGW, WLR}: this was the first direct detection. Several detections immediately followed opening the doors to the {\it prova regina} for any theory of gravity: features like gravitational wave polarization, amplitude, spin and so on can single out the underlying theory of gravity. 

Also ETGs predict the existence of gravitational waves by solving the associated linearized field equations. However different properties with respect to General Relativity can emerge. While the Einstein gravitational radiation is quadrupolar, transverse and with helicity 2 - carried by massless spin 2 graviton - the one related to  ETGs can have  transverse and longitudinal polarizations with 0 and 2 helicity,  carried by massive and massless  gravitons with spin  0 and 2.  As pointed out in literature \cite{BCDN}, also ghost modes can emerge  into dynamics. Here, we are going to investigate the weak field limit of higher order gravity and the related gravitational radiation putting in evidence the main differences with respect to General Relativity. 

The paper is organized in the following way: in Sec. 2 we derive the field equations for higher-order Lagrangian including  terms like $R\Box^{k}R$ by using a variational principle and then we derive the  weak field limit. In Sec. 3, we recover the Newtonian limit  reproducing the Poisson equation as a consistency check. Sec. 4 is devoted to the  resolution of  the linear field equations by using the Fourier transformation and the  Green functions. These propagators give rise to the gravitational waves. In Sec. 5, the gravitational waves  are analyzed by  the polarization and helicity of the massive and non-massive modes. Conclusions are drawn in Sec.6.

\section{The field equations and  the weak field limit}
Let us consider the Lagrangian density $L$ and its variational derivative with respect to $g_{\mu\nu}$
\begin{equation}
\label{laglag}
\begin{aligned}
    & L=L_{g}+L_{m}\ , \\
    & L_{g}=(R+a_{0}R^{2}+\sum_{k=1}^{p} a_{k}R\Box^{k}R)\sqrt{-g}\ ,
 &P^{\mu\nu}=-\frac{1}{\sqrt{-g}} \frac{\delta L_{g}}{\delta g_{\mu\nu}}\ ,\\
    & L_{m}=2\chi\sqrt{-g}\mathcal{L}_{m}\ , \ \ &T^{\mu\nu}=\frac{2}{\sqrt{-g}}\frac{\delta \left(\sqrt{-g}\mathcal{L}_{m}\right)}{\delta g_{\mu\nu}}\ ,
\end{aligned} 
\end{equation}
with $\chi={8\pi G}/{c^{4}}$ and $\Box=g^{\mu\nu}\nabla_{\mu}\nabla_{\nu}$ the covariant D'Alembert operator. Here $P^{\mu\nu}$ is the tensor as defined in \cite{CCT1} relative to the gravitational Lagrangian and $T^{\mu\nu}$ is the matter energy-momentum tensor.  Applying  the principle of minimal  action with the stationarity condition  $\delta L/\delta g_{\mu\nu}=0$, we can define \cite{Sixt order, Var derivative}:
\begin{equation}
\label{lag}
\frac{\delta L}{\delta g_{\mu\nu}}=\sum_{m=0}^{2p+2}\left(-1\right)^{m}\left(\frac{\partial L}{\partial g_{\mu\nu,i_{1}\cdots i_{m}}}\right)_{,i_{1}\cdots i_{m}} \ .
\end{equation}
We then obtain the field equations related to the above Lagrangian density \footnote{The signature of the metric $g_{\mu\nu}$ is $(+\ \ , -\ \ , -\ \ , -)$, the Ricci tensor is defined as
$R_{\mu\nu}=R_{\ \ \mu\rho\nu}^{\rho}$ and the Riemann tensor as $R_{\ \ \beta\mu\nu}^{\alpha}=\Gamma_{\beta\nu,\mu}^{\alpha}$+\ldots.} $L$:
\begin{equation}
\begin{split}
{\cal G}^{\mu\nu}\equiv R^{\mu\nu}-\frac{1}{2}g^{\mu\nu}R+2a_{0}(RR^{\mu\nu}-\frac{1}{4}R^{2}g^{\mu\nu}-R^{;\mu\nu}+g^{\mu\nu}\Box R)\\ +2\sum_{j=1}^{p}a_{j}(R^{\mu\nu}\Box^{j}R-\frac{1}{4}g^{\mu\nu}R\Box^{j}R-(\Box^{j}R)^{;\mu\nu}+g^{\mu\nu}\Box^{j+1}R) \\
+\sum_{A=1}^{p}\sum_{j=A}^{p}a_{j}\frac{1}{2}\biggl \{ g^{\mu\nu}\biggl [\left(\Box^{j-A}R\right)_{;\lambda}\left(\Box^{A-1}R\right)^{;\lambda}+\left(\Box^{j-A}R\right)\left(\Box^{A}R\right)\biggr ]\\-2(\Box^{j-A}R)^{;\mu}(\Box^{A-1}R)^{;\nu} \biggr \}=\chi T^{\mu\nu} \ .
\end{split}
\end{equation}
In the weak field limit, $g_{\mu\nu}=\eta_{\mu\nu}+h_{\mu\nu}$ with $|{h_{\mu\nu}}| \ll1$, the field equations became \cite{Weinb, CARLN, CST}:
\begin{multline}\label{eq prem}
\frac{1}{2} (h_{\ \mu,\nu\rho}^{\rho}+h_{\ \nu,\mu\rho}^{\rho}-\Box h_{\mu\nu}-h_{,\mu\nu})-\left(\eta_{\mu\nu}\eta_{\lambda\sigma}+2\eta_{\mu\lambda}\eta_{\nu\sigma}\right)\sum_{k=0}^{p} a_{k} \Box^{k}(h_{\ \ ,\tau\kappa}^{\tau\kappa}-\Box h)^{,\lambda\sigma}\\=4\pi (2 \ T_{\mu\nu}-T\eta_{\mu\nu})\ ,
\end{multline}
with $\Box=\eta^{\mu\nu}\partial_{\mu}\partial_{\nu}$. To verify the contracted Bianchi identities $\nabla_{\nu}{\cal G}^{\mu\nu}=0$, the energy-momentum tensor $T_{\mu\nu}$ must be unperturbed that is approximated to zero-order in $h$, so that the conservation law is not violated, $\partial^{\nu}T_{\mu\nu}=0$. In fact the ordinary divergence of left- and right-hand  sides of (\ref{eq prem}) must be the same, that is:
\begin{multline}\label{bianchiverif}
\frac{1}{2} (h_{\ \mu,\nu\rho}^{\rho\ ,\mu}+\cancel{h_{\ \nu,\mu\rho}^{\rho\ ,\mu}}-\cancel{\Box h_{\mu\nu}^{\ \ ,\mu}}-h_{,\mu\nu}^{,\mu})-\left(\eta_{\mu\nu}\eta_{\lambda\sigma}+2\eta_{\mu\lambda}\eta_{\nu\sigma}\right)\sum_{k=0}^{p} a_{k} \Box^{k}(h_{\ \ ,\tau\kappa}^{\tau\kappa}-\Box h)^{,\lambda\sigma\mu}\\=4\pi (2 \ T_{\mu\nu}^{\ \ ,\mu}-T^{,\mu}\eta_{\mu\nu}) \ .
\end{multline}
The trace of (\ref{eq prem}) gives
\begin{equation}\label{trac1}
6\sum_{k=0}^{p} a_{k} \Box^{k}\left(h_{\ \ ,\tau\kappa}^{\tau\kappa}-\Box h\right)=h_{\ \ ,\tau\kappa}^{\tau\kappa}-\Box h +8\pi T \ ,
\end{equation}
and, from Eqs. (\ref{bianchiverif}) and (\ref{trac1}),  we have:
\begin{equation}
\frac{1}{2}\left(h_{\ \ ,\tau\kappa}^{\tau\kappa}-\Box h\right)_{,\nu}-\frac{1}{6}\left(\eta_{\mu\nu}\eta_{\lambda\sigma}+2\eta_{\mu\lambda}\eta_{\nu\sigma}\right)\left(h_{\ \ ,\tau\kappa}^{\tau\kappa}-\Box h +8\pi T\right)^{,\lambda\sigma\mu}=4\pi \left(2 \ T_{\mu\nu}^{\ \ ,\mu}-T^{,\mu}\eta_{\mu\nu}\right)\,,
\end{equation}
namely:
\begin{equation}
\frac{1}{2}\left(h_{\ \ ,\tau\kappa}^{\tau\kappa}-\Box h\right)_{,\nu}-\frac{1}{6}\left(3\Box\right)\left(h_{\ \ ,\tau\kappa}^{\tau\kappa}-\Box h +8\pi T\right)_{,\nu}=4\pi \left(2 \ T_{\mu\nu}^{\ \ ,\mu}-T_{,\nu}\right) \ ,
\end{equation}
verified for $\partial^{\nu}T_{\mu\nu}=0$.\\
Eq. (\ref{eq prem}) is gauge-invariant that is a gauge transformation, such that the harmonic gauge is  always verified, exists. It is 
\begin{equation}
(h^{\mu\nu}-\frac{1}{2} \eta^{\mu\nu}h)_{,\mu}=0 \Rightarrow
\begin{cases}
h_{\ \ ,\tau\kappa}^{\tau\kappa}=\frac{1}{2}\Box h \\ h_{\ \mu,\nu\rho}^{\rho}+h_{\ \nu,\mu\rho}^{\rho}-h_{,\mu\nu}=0
\end{cases}\ .
\end{equation}
The simplified equations gives:
\begin{equation}
\label{eq pm}
\boxed{\Box h_{\mu\nu}-(\eta_{\mu\nu}\eta_{\lambda\sigma}+2\eta_{\mu\lambda}\eta_{\nu\sigma})\sum_{k=0}^{p} a_{k} \Box^{k+1} h^{,\lambda\sigma}=-8\pi(2\ T_{\mu\nu}^{\left(0\right)}-T^{\left(0\right)}\eta_{\mu\nu})} \ ,
\end{equation}
namely ten linear $2(p+2)$-order  partial differential equations where six of them are independent because of the contracted Bianchi identities. The trace of Eq. \eqref{eq pm} is:
\begin{equation}
\label{traccia}
\Box h-6\sum_{k=0}^{p}a_{k}\Box^{k+2}h=16\pi T \ .
\end{equation}
In more compact form, defining
\begin{equation}
\begin{cases}
c_{1}=1 \\ c_{l}=-6a_{l-2} \quad  \text{if $l>1$} \ ,
\end{cases}
\end{equation}
we get then 
\begin{equation}
\label{traccia def}
\boxed{\sum_{l=1}^{p+2} c_{l} \Box^{l} h=16\pi T^{\left(0\right)}} 
\end{equation}
which is one of the main results of the present paper.  It means that higher-order terms in the $\Box$ operator contribute to the gravitational wave equation as further derivatives. As we will see below, this fact is important in the Fourier analysis of the waves.

\section{The Newtonian limit}
Let us now consider the Newtonian approximation assuming that, beside the weak field, also the involved velocities are less that the speed of light. We will consider two approaches to deal with the Newtonian limit.
\subsection{First approach}
In the Newtonian approximation we impose that:
\begin{enumerate}
\item the field is weak $g_{\mu\nu}=\eta_{\mu\nu}+h_{\mu\nu}$ with $|{h_{\mu\nu}}| \ll 1$;
\item the field is static $g_{\mu\nu,0}=0$ and $g_{i0}=0$;
\item the velocities are small compared to $c$ that is  $v/c\ll1$ (slow motion)\,.
\end{enumerate}
Using the matter energy-momentum tensor $T_{\mu\nu}$ at the zero order,   $T_{\mu\nu}=\delta_{\mu}^{0}\delta_{\nu}^{0}\rho$, because we are in the slow motion approximation, from \eqref{eq pm} we get the following field equations \cite{Limit New, NLFO, HSJ}:
\begin{equation}
\label{limit newt}
\Delta h_{\mu\nu}+(\eta_{\mu\nu}\eta_{\lambda\sigma}+2\eta_{\mu\lambda}\eta_{\nu\sigma})\sum_{k=0}^{p}a_{k}(-1)^{k+1}\Delta^{k+1} h^{,\lambda\sigma}=8\pi(2\delta_{\mu}^{0}\delta_{\nu}^{0}-\eta_{\mu\nu})\rho \ ,
\end{equation}
whose trace is:
\begin{equation}
\label{traccia limit newt}
\Delta h+6\sum_{k=0}^{p}a_{k}(-1)^{k+2}\Delta^{k+2} h=-16\pi\rho \ .
\end{equation}
Here $\rho$ is the perfect-fluid matter density.
From \eqref{limit newt}, the equation of  $time-time$ component is:
\begin{equation}
\label{time time}
\Delta h_{00}+\sum_{k=0}^{p}a_{k}(-1)^{k+2}\Delta^{k+2} h=+8\pi\rho \ .
\end{equation}
From \eqref{limit newt} the equations of the space components are:
\begin{equation}
\Delta h_{ij}+\eta_{ij}\sum_{k=0}^{p}a_{k}(-1)^{k+2}\Delta^{k+2} h+2\sum_{k=0}^{p}(-1)^{k+1}\Delta^{k+1}h_{,ij}=8\pi\rho\delta_{ij} \ .
\end{equation}
From \eqref{traccia limit newt} and \eqref{time time} we obtain:
\begin{equation}
\label{time time 1}
6\Delta h_{00}-\Delta h=64\pi\rho \ .
\end{equation}
The metric perturbation $h_{\mu\nu}$ in harmonic gauge is
\begin{equation}
\label{gauge}
   h_{\ \ ,j}^{ij}-\frac{1}{2}h^{,i}=0 \ ,
\end{equation}
which is:
\begin{equation}
\label{system gauge}
\begin{cases}
h_{\ \ ,1}^{11}+h_{\ \ ,2}^{12}+h_{\ \ ,3}^{13}+\frac{1}{2}h_{,1}=0 \\
h_{\ \ ,1}^{21}+h_{\ \ ,2}^{22}+h_{\ \ ,3}^{23}+\frac{1}{2}h_{,2}=0 \\ 
h_{\ \ ,1}^{31}+h_{\ \ ,2}^{32}+h_{\ \ ,3}^{33}+\frac{1}{2}h_{,3}=0 \ ,
\end{cases}
\end{equation}
verified if:
\begin{equation}
\label{comp non null gauge}
h_{00}=h_{11}=h_{22}=h_{33} \ ,
\end{equation}
and the remaining ones equal to zero.
From \eqref{time time 1} and  \eqref{comp non null gauge}, knowing that $h=-2h_{00}$, we get:
\begin{equation}
\label{eq poisson h0}
\Delta h_{00}=8\pi\rho \ .
\end{equation}
Setting $h_{00}=2\phi$ we have:
\begin{equation}
h_{\mu\nu}=\text{diag}(2\phi,2\phi,2\phi,2\phi) \ .
\end{equation}
Substituting in \eqref{eq poisson h0} we obtain the standard Poisson equation
\begin{equation}
\Delta \phi=4\pi\rho \ .
\end{equation}
This means that the Newtonian limit of the theory is consistent with the one of General Relativity.
\subsection{Second approach}
The Poisson equation can be obtained in an alternative way by adding to the metric the  hypotheses of asymptotic flatness and asymptotic behavior. We can relax the hypothesis of the static spacetime that becomes stationary \cite{LL, SC, RINDLER}. Considering all the hypotheses, we have
\begin{enumerate}
\item the weak field limit; 
\item the stationarity ($g_{\mu\nu,0}=0$);
\item the slow motion;
\item the asymptotic flatness;
\item the asymptotic behavior.
\end{enumerate}
Performing the following replacement:
\begin{equation}
\bar{h}^{\mu\nu}=h^{\mu\nu}-\frac{1}{2}\eta^{\mu\nu}h \ ,
\end{equation}
the harmonic gauge becomes $(\bar{h}^{\mu\nu})_{,\mu}=0$.
If we impose that the metric is:
\begin{enumerate}
\item
\textit{stationarity} 
\begin{equation*} 
g_{\mu\nu,0}=0\ ,
\end{equation*}
\item
\textit{asymptotically flat}  
\begin{equation*}
\lim_{r \to \infty}g_{\mu\nu}=\eta_{\mu\nu}\ ,
\end{equation*}
\item
\textit{asymptotic behavior} 
\begin{equation*}
\lim_{r\to\infty}r^{M}\left[\bar{h}^{\mu\nu}-\sum_{n=0}^{M}\frac{A_{n}^{\mu\nu}}{r^{n}}\right]=0 \quad \forall M>0 \ ,
\end{equation*}
\end{enumerate}
we can asymptotically develop the tensor   $\bar{h}^{\mu\nu}$ in the following way:
\begin{equation}
\label{sviluppo h}
\bar{h}^{\mu\nu}=\frac{A^{\mu\nu}}{r}+\mathcal{O}\left(\frac{1}{r^2}\right) \ .
\end{equation}
Imposing the gauge condition to \eqref{sviluppo h}:
\begin{equation}
0=\bar{h}_{\ \ ,\nu}^{\mu\nu}=\bar{h}_{\ \ ,i}^{\mu i}=\frac{A^{\mu i}n_{i}}{r^2}+\mathcal{O}\left(\frac{1}{r^{3}}\right) \quad \text{with} \quad n_{i}=\frac{x_{i}}{r} \ ,
\end{equation}
we get:
\begin{equation}
A^{\mu i}=0, \qquad \lvert \bar{h}^{00} \rvert \gg\lvert \bar{h}^{0i}\rvert, \qquad  \lvert \bar{h}^{00} \rvert \gg\lvert \bar{h}^{ki}\rvert \ ,
\end{equation}
that is,  only  $\bar{h}^{00}$ survives to the order $\frac{1}{r}$ because $\bar{h}^{\mu i}\approx \frac{1}{r^{2}}$.
In general,  we have:
\begin{equation}
\Delta^{k+1}\bar{h}^{,\lambda\sigma} \approx \frac{1}{r^{2k+5}} \quad \text{and} \quad \Delta \bar{h}_{\mu\nu} \approx \frac{1}{r^3} \quad \Rightarrow \Delta^{k+1}\bar{h}^{,\lambda\sigma}\ll\Delta \bar{h}_{\mu\nu} \ .
\end{equation}
So Eq. \eqref{limit newt}, in these hypotheses (weak field, asymptotic flatness and asymptotic behavior), becomes 
\begin{equation}
\Delta \bar{h}_{00}=16\pi \left(2T_{00}-T\right) \ ,
\end{equation}
because $\bar{h}_{00}=2h_{00}$.\\
Considering the further hypothesis that  velocities are much lower than $c$,  i.e. the \textit{slow motion}, we get:
\begin{equation}
\left| \frac{v}{c} \right| \ll1\rightarrow \lvert T^{00} \rvert \gg \lvert T^{0i} \rvert \gg \lvert T^{ij} \rvert \rightarrow T=T_{00}=\rho \ ,
\end{equation}
and then $$
\Delta {h}_{00}=8\pi\rho \ .
$$
Setting $h_{00}=2\phi$ the Poisson equation is again obtained
$$
\Delta \phi=4\pi\rho \ .
$$
It is worth saying that the above approaches are equivalent from a practical point of view.  However, the energetic behavior of the gravitational field is better stressed in the second  than in the first one due to the considerations on the asymptotic behavior. In the first case, the structure of the weak filed metric is better defined. 

\section{The solution of the field equations}

Let us  solve the differential equations with the Fourier transform method on an unlimited domain $R^4$ with the hypothesis of sufficient regularity for $T_{\mu\nu}\in L^{2}\left(R^{4}\right)$. 
In order to integrate Eqs. \eqref {eq pm}, we first solve its trace by calculating $h(x)$ from  \eqref {traccia def}, replace it in the field equations which, at this point,  become a non-homogeneous wave-like equation and finally, we obtain  the metric perturbation $h_{\mu\nu}$. We write the general integral as the sum of the homogeneous solution plus a particular solution due to the linearity of Eqs.\eqref{eq pm} that is
\begin{equation}
h_{\mu\nu}=h_{\mu\nu}^{homog}+h_{\mu\nu}^{part}\ ,
\end{equation}
where we indicate  $\mathfrak{h}_{\mu\nu}$ and $\mathtt{h}_{\mu\nu}$ as the homogeneous and the particular solution respectively.
\subsection{The particular solution}
Let us calculate the Green function for the linear differential operator
\begin{equation}
D=\sum_{l=1}^{p+2} c_{l} \Box^{l}\ ,
\end{equation}
namely
\begin{equation}
\label{Green Box}
\sum_{l=1}^{p+2} c_{l} \Box^{l} G_{D}(x,x')=\delta^{4}(x-x')\ .
\end{equation}
Fourier integrals for the $G_{D}$ and $\delta$ are
\begin{equation}
G_{D}(x,x')=\int \frac{d^{4}k}{(2\pi )^{4}} \tilde{G}_{D}(k)e^{ik^{\alpha}(x_{\alpha}-x_{\alpha}')}\ , \quad
\delta(x-x')=\int \frac{d^{4}k}{(2\pi )^{4}}e^{ik^{\alpha}(x_{\alpha}-x_{\alpha}')}\ ,
\end{equation}
where $x=x^{\alpha}=(t,\bf{x})$, $k=k^{\alpha}=(\omega,\bf{k})$, $k^{2}=k^{\alpha}k_{\alpha}$, $\mathbf{k} \cdot \mathbf{k} = q^{2}$ and $q=\left|\mathbf{k}\right|$.\\
By means of the formula
\begin{equation}
\Box^{l}G_{D}(x,x')=\int \frac{d^{4}k}{(2\pi )^{4}} \left(-1\right)^{l}k^{2l}\tilde{G}_{D}(k)e^{ik^{\alpha}(x_{\alpha}-x_{\alpha}')}\ ,
\end{equation}
writing Eq. \eqref{Green Box} in the space of $k$, we get the  transformed Green function
\begin{equation}
 \tilde{G}_{D}(k)=\frac{1}{\sum_{l=1}^{p+2} c_{l} (-1)^{l} k^{2l}}\ ,
\end{equation}
that is
\begin{equation}
G_{D}(x,x')=\int \frac{d^{4}k}{(2\pi )^{4}} \frac{1}{\sum_{l=1}^{p+2} c_{l}(-1)^{l}k^{2l}} e^{ik^{\alpha}(x_{\alpha}-x_{\alpha}')}\ .
\end{equation}
So the particular solution of \eqref{traccia def} is:
\begin{equation}
\mathtt{h}(x)=16\pi\int d^{4}x' G_{D}(x,x')T(x')\ ,
\end{equation}
namely
\begin{equation}
\label{sol par traccia}
\mathtt{h}(x)=\int \frac{d^{4}k}{(2\pi)^{4}}\underbrace{\left[ \int d^{4}x' \frac{16\pi T(x')e^{-ik^{\alpha}x_{\alpha}'}}{\sum_{l=1}^{p+2} c_{l}(-1)^{l}k^{2l}}\right]}_{\tilde{\mathtt{h}}\left(k\right)}e^{ik^{\alpha}x_{\alpha}}\ .
\end{equation}
Each derivative of the metric perturbation $h_{\mu\nu}$,  in the coordinate space, adds one $ik_{\mu}$ in the Fourier space, that is $h\left(x\right)_{,\mu} \rightarrow ik_{\mu}\tilde{h}\left(k\right)$. Replacing Eq. \eqref{sol par traccia} into  \eqref{eq pm},  we get
\begin{equation}
\Box \mathtt{h}_{\mu\nu}-\underbrace{\int \frac{d^{4}k}{(2\pi)^{4}}\left\{\tilde{\mathtt{h}}(k)\left[\left(\eta_{\mu\nu}k^{2}+2k_{\mu}k_{\nu}\right)\sum_{l=0}^{p}a_{l}(-1)^{l+2}k^{2(l+1 )}\right]e^{ik^{\alpha}x_{\alpha}}\right\}}_{F_{\mu\nu}(x)}=-8\pi(2T_{\mu\nu}-T\eta_{\mu\nu})\ ,
\end{equation}
and thus we have
\begin{equation}
\Box \mathtt{h}_{\mu\nu}=-8\pi(2\ T_{\mu\nu}-T\eta_{\mu\nu})+F_{\mu\nu}(x)\ ,
\end{equation}
\begin{equation}
\boxed{\mathtt{h}_{\mu\nu}\left(x\right)=\int d^{4}x''G_{\Box}(x,x'')\left[-8\pi(2\ T_{\mu\nu}(x'')-\eta_{\mu\nu}T(x''))+F_{\mu\nu}(x'')\right]}\ ,
\end{equation}
namely the \emph{particular solution of the field equations}, where $G_{\Box}$ is the Green function for the $\Box$ operator:
\begin{equation}
\Box G_{\Box}(x,x'')=\delta^{4}(x-x'')\ ,
\end{equation}
\begin{equation}
G_{\Box}(x,x'')=\int\frac{d^{4}k}{(2\pi)^{4}}\left(-\frac{1}{k^{2}-i\epsilon}\right)e^{ik^{\alpha}(x_{\alpha}-x''_{\alpha})}\ .
\end{equation}
From a physical point of view, this solution is interesting because, due to the presence of $T_{\mu\nu}$ in it,  it is strictly related to the distribution of matter-energy which determines the propagation of the gravitational interaction.

\subsection{The homogeneous solution}
In order to calculate the homogeneous solution of Eqs. \eqref{eq pm}, we perform the Fourier transform only of the spatial coordinates of the trace:
\begin{equation}
\mathfrak{h}\left(t,\bf{x}\right)=\int\frac{d^{3}\bf{k}}{\left(2\pi\right)^{3}}\mathfrak{h}\left(t,\bf{k}\right)e^{-i\bf{k} \cdot \bf{x}}\ ,
\end{equation}
by means of the homogeneous Eq. \eqref{traccia def} 
\begin{equation}
\sum_{l=1}^{p+2}c_{l}\left( \partial_{0}^{2}-\Delta \right)^{l}\mathfrak{h}(x)=0\ ,
\end{equation}
Then we obtain:
\begin{equation}
\sum_{l=1}^{p+2}c_{l}\left[\frac{\partial^{2}}{\partial t^{2}}+q^{2}\right]^{l}\mathfrak{h}(t,\mathbf{k})=0\ ,
\end{equation}
where $q^{2}=\mathbf{k}\cdot\mathbf{k}$.
Let $\tilde{D}_{t}$ be the following linear differential operator
$$
\tilde{D}_{t}=\partial_{0}^{2}+q^{2}\ ,
$$
we rewrite the equation in the following way
\begin{equation}
\label{eq diff traccia}
\sum_{l=1}^{p+2}c_{l}\tilde{D_{t}}^{l}\mathfrak{h}(t,\mathbf{k})=0\ ,
\end{equation}
representing a homogeneous ordinary differential equation in $t$ of  $2(p + 2)$-degree, with constant coefficients. Its characteristic polynomial is:
\begin{equation}
\label{pol car}
\sum_{l=1}^{p+2}c_{l}\left( \lambda^{2}+q^{2}\right)^{l}=0\ ,
\end{equation}
which admits $2(p + 2)$ solutions in $\lambda$, for the fundamental theorem of the algebra. If we set
\begin{equation}
\lambda_{m}^{\pm}=\pm i\omega_{m}\ ,
\end{equation}
we obtain $(p+2)$  polynomial solutions for $\omega_{m}$. We define the four-vector $k_{m}^{\mu}=\left(\omega_{m},\mathbf{k}\right)$ with
\begin{equation}
k_{m}^{2}=\omega_{m}^{2}-q^{2}\ ,
\end{equation}
and we have $(p+2)$ solutions $k_{m}^{2}$ of the equation $\sum_{l=1}^{p+2}c_{l}\left(-1\right)^{l}\left(k^{2}\right)^{l}=0$. 
A solution is definitely $k_{1}^{2}=0$, i.e. $\omega_{1}=q$, that  give the gravitational waves predicted by General Relativity. In general $\omega_{m}$  is a complex number and therefore there are also  damped waves, but, considering  only real $\omega_{m}$, one can   impose conditions on $c_{l}$,  that is on $a_{l}$. So we will limit  to  the case of real, distinct $\omega_{m}$: this means that they are exactly $(p+2)$ different real numbers  from the imposed conditions. The solutions of Eq. \eqref{eq diff traccia} are  \cite{LL, MTW}:
\begin{equation}
\mathfrak{h}\left(t,\mathbf{k}\right)=\sum_{m=1}^{p+2}\left[A'_{m}\left(\mathbf{k}\right) e^{i\omega_{m}t}+A''_{m}\left(\mathbf{k}\right) e^{-i\omega_{m}t}\right]\ ,
\end{equation}
which, in  coordinate space, becomes:
\begin{equation}
\mathfrak{h}\left(t,\mathbf{x}\right)=\sum_{m=1}^{p+2}\int \frac{d^{3} \mathbf{k}}{(2\pi)^{3}}\left[A_{m}^{+}\left(\mathbf{k}\right) e^{i\left(\omega_{m}t- \mathbf{k}\cdot\mathbf{x}\right)}+A_{m}^{-}\left(\mathbf{k}\right) e^{-i\left(\omega_{m}t- \mathbf{k}\cdot\mathbf{x}\right)}\right]\ ,
\end{equation}
where $A_{m}^{+}\left(\mathbf{k}\right)=A'_{m}\left(\mathbf{k}\right)$ and $A_{m}^{-}\left(\mathbf{k}\right)=A''_{m}\left(\mathbf{-k}\right)$.
Since we want a real $\mathfrak{h}\left(t,\mathbf{x}\right)$  so that $A_{m}^{+}\left(\mathbf{k}\right)=\bar{A}_{m}^{-}\left(\mathbf{k}\right)$ , we take the real part of the following integral:
\begin{equation}
\label{modi fou traccia}
\mathfrak{h}\left(t,\mathbf{x}\right)=\Re \left\{\sum_{m=1}^{p+2}\int \frac{d^{3} \mathbf{k}}{(2\pi)^{3}}A_{m}\left(\mathbf{k}\right) e^{i\left(\omega_{m}t- \mathbf{k}\cdot\mathbf{x}\right)}\right\}\ ,
\end{equation}
where $A_{m}$ is a complex function. Later, even if not indicated, we will consider only  the real part or equivalently the c.c..
Let us represent $\mathfrak{h}_{\mu\nu}\left(t,\mathbf{x}\right)$ by means of the Fourier transform with respect to the spatial coordinates only $\mathbf{x}$
\begin{equation}
\label{modi fou perturb}
\mathfrak{h}_{\mu\nu}\left(t,\mathbf{x})\right)=\int \frac{d^{3}\mathbf{k}}{\left(2\pi\right)^{3}}\mathfrak{h}_{\mu\nu}\left(t,\mathbf{k}\right)e^{-i \mathbf{k}\cdot\mathbf{x}}\ ,
\end{equation}
where $k_{m}^{\mu}=(\omega_{m},\mathbf{k})$ as already defined. From the homogeneous Eq. \eqref{eq pm}, from the \eqref{modi fou traccia} and \eqref{modi fou perturb} in the Fourier space,  we have:
\begin{equation}
\label{eq diff in k}
\left(\partial_{0}^{2}+q^{2}\right)\mathfrak{h}_{\mu\nu}\left(t,\mathbf{k}\right)=\sum_{m=1}^{p+2}\left\{\left[\eta_{\mu\nu}k_{m}^{2}+2\left(k_{m}\right)_{\mu}\left(k_{m}\right)_{\nu}\right]\sum_{l=0}^{p}a_{l}\left(-1\right)^{l+2}k_{m}^{2\left(l+1\right)}\right\}A_{m}\left(\mathbf{k}\right) e^{i\omega_{m}t}\ ,
\end{equation}
where $k_{m}^{2}=\left(k_{m}\right)^{\alpha}\left(k_{m}\right)_{\alpha}=\omega_{m}^{2}-q^{2}$.\\
For $m=1$, being $k_{1}^{2}=0$, the first term of the sum in $m$ in Eq. \eqref{eq diff in k} is cancelled  and then the sum starts from $m=2$.
For  $2\leq m \leq p+2$, where $k_{m}^{2}\neq0$, from the following identity:
\begin{equation}
\sum_{l=1}^{p+2}c_{l}\left(-1\right)^{l}k_{m}^{2l}=-k_{m}^{2}\left(1+6\sum_{l=0}^{p}a_{l}\left(-1\right)^{l+2}k_{m}^{2\left(l+1\right)}\right)=0\Rightarrow \sum_{l=0}^{p}a_{l}\left(-1\right)^{l+2}k_{m}^{2\left(l+1\right)}=-\frac{1}{6}\ ,
\end{equation}
we get:
\begin{equation}
\left(\partial_{0}^{2}+q^{2}\right)\mathfrak{h}_{\mu\nu}\left(t,\mathbf{k}\right)=\sum_{m=2}^{p+2}\rho_{\mu\nu}\left(\mathbf{k};m\right)e^{i\omega_{m}t}\ ,
\end{equation}
where
\begin{equation}
\rho_{\mu\nu}\left(\mathbf{k};m\right)=\left\{-\frac{k_{m}^{2}}{3}\left[\frac{\eta_{\mu\nu}}{2}+\frac{\left(k_{m}\right)_{\mu}\left(k_{m}\right)_{\nu}}{k_{m}^{2}}\right]\right\}A_{m}\left(\mathbf{k}\right)\ .
\end{equation}
\\
The Fourier transform in the $\mathbf{x}$ coordinates of the function $\rho_{\mu\nu}\left(t,\mathbf{x}\right)$ is
\begin{equation}
    \rho_{\mu\nu}\left(x\right)=\sum_{m=2}^{p+2}\int\frac{d^{3}\mathbf{k}}{(2\pi)^{3}}\rho_{\mu\nu}\left(\mathbf{k};m\right)e^{ik_{m}^{\alpha}x_{\alpha}}=\sum_{m=2}^{p+2}\rho_{\mu\nu}\left(x;m\right)\ ,
\end{equation}
where $k_{m}^{\alpha}x_{\alpha}=\omega_{m}t-\mathbf{k}\cdot\mathbf{x}$. In the space of $x$, we obtain
\begin{equation}
 \Box \mathfrak{h}_{\mu\nu}\left(x\right)=\sum_{m=2}^{p+2}\rho_{\mu\nu}\left(x;m\right)\ ,
\end{equation}
that we solve as the sum of the homogeneous solution plus the particular one:
\begin{equation}
  \mathfrak{h}_{\mu\nu}\left(x\right)=\underbrace{\int\frac{d^{3}\mathbf{k}}{\left(2\pi\right)^{3}}C_{\mu\nu}\left(\mathbf{k}\right)e^{ik_{1}^{\alpha}x_{\alpha}}}_{\text{classical gravitational waves in vacuum}}+\underbrace{\sum_{m=2}^{p+2}\int d^{4}x'G_{\Box}\left(x,x'\right)\rho_{\mu\nu}\left(x';m\right)}_{\text{correction  terms}}\ ,
\end{equation}
that is the \emph{solution of the field equations in the vacuum}. Equivalently:
\begin{equation}
\label{sol vuoto eq}
\mathfrak{h}_{\mu\nu}\left(x\right)=\underbrace{\int\frac{d^{3}\mathbf{k}}{\left(2\pi\right)^{3}}C_{\mu\nu}\left(\mathbf{k}\right)e^{ik_{1}^{\alpha}x_{\alpha}}}_{\text{classical gravitational waves in vacuum}}+\underbrace{\sum_{m=2}^{p+2}\int\frac{d^{3}\mathbf{k}}{\left(2\pi\right)^{3}}G_{\Box}\left(k_{m}\right)\rho_{\mu\nu}\left(\mathbf{k};m\right)e^{ik_{m}^{\alpha}x_{\alpha}}}_{\text{correction terms}}\ ,
\end{equation}
that we can also write more explicitly
\begin{equation}
\label{sol vuoto eq1}
\boxed{
\mathfrak{h}_{\mu\nu}\left(x\right)=\underbrace{\int\frac{d^{3}\mathbf{k}}{\left(2\pi\right)^{3}}C_{\mu\nu}\left(\mathbf{k}\right)e^{ik_{1}^{\alpha}x_{\alpha}}}_{\text{classical gravitational waves in vacuum}}+\underbrace{\sum_{m=2}^{p+2}\int\frac{d^{3}\mathbf{k}}{\left(2\pi\right)^{3}}\left\{\frac{1}{3}\left[\frac{\eta_{\mu\nu}}{2}+\frac{\left(k_{m}\right)_{\mu}\left(k_{m}\right)_{\nu}}{k_{m}^{2}}\right]\right\}A_{m}\left(\mathbf{k}\right)e^{ik_{m}^{\alpha}x_{\alpha}}}_{\text{correction terms}}
}\ .
\end{equation}
So the \emph{general integral of the field equations in presence of matter} is:
\begin{equation}
\boxed{
\begin{aligned}
 h_{\mu\nu}\left(x\right)&=\underbrace{\int d^{4}x'G_{\Box}(x,x')\left[-8\pi(2\ T_{\mu\nu}(x')-\eta_{\mu\nu}T(x'))\right]+\int\frac{d^{3}\mathbf{k}}{\left(2\pi\right)^{3}}C_{\mu\nu}\left(\mathbf{k}\right)e^{ik^{\alpha}x_{\alpha}}}_{\text{gravitational waves in matter}}\\
  &+\underbrace{\int d^{4}x'G_{\Box}(x,x')\left[F_{\mu\nu}\left(x'\right)+\sum_{m=2}^{p+2}\rho_{\mu\nu}\left(x';m\right)\right]}_{\text{correction terms}}
\end{aligned}
}\ .
\end{equation}
This is the most general gravitational wave solution coming from  higher-order gravitational theories.

\section{Polarization and  helicity states in  vacuum}
In order to study the polarization and the helicitity of waves \cite{Wave 4a, STELLE, BCDN}, let us consider separately the mode with $k_{1}^{2}=0$, that we indicate with the oscillation mode A$_{1}$, from the $(p+1)$ modes with $k_{m}^{2}\ne0$ for $2\leq m \leq p+2$, that we indicate with the A$_{m}$ modes. All the modes are  classified in  Table \ref{classificazione}.
\begin{table}[H]
\[\footnotesize{
\begin{array}{clc}
\toprule
\text{Mode}  & \text{Wave number} & \text{Gauge condition}\\
\midrule
\text{A}_{1}  & k^{2}=k_{1}^{2}=0 \wedge \tilde{\mathfrak{h}}=C_{\mu}^{\ \mu}\left(\mathbf{k}\right)=0 \Rightarrow \text{any} \, \tilde{\mathfrak{h}}_{\mu\nu} & \tilde{\mathfrak{h}}_{\mu\nu}k^{\mu}=0 \\
\midrule
\text{A}_{2} & k^{2}=k_{2}^{2}=\omega_{2}^{2}-q^{2} \wedge \tilde{\mathfrak{h}}=A_{2}\left(\mathbf{k}\right)\neq0 \Rightarrow \tilde{\mathfrak{h}}_{\mu\nu}=\frac{1}{3}\left(\frac{\eta_{\mu\nu}}{2}+\frac{\left(k_{2}\right)_{\mu} \left(k_{2}\right)_{\nu}}{k_{2}^{2}}\right)\tilde{\mathfrak{h}} & \tilde{\mathfrak{h}}_{\mu\nu}k^{\mu}-\frac{1}{2}\tilde{\mathfrak{h}}k_{\nu}=0 \\
\midrule
\vdots & \vdots & \vdots\\
\midrule
\text{A}_{p+2} & k^{2}=k_{p+2}^{2}=\omega_{p+2}^{2}-q^{2} \wedge \tilde{\mathfrak{h}}=A_{m}\left(\mathbf{k}\right)\neq0 \Rightarrow \tilde{\mathfrak{h}}_{\mu\nu}=\frac{1}{3}\left(\frac{\eta_{\mu\nu}}{2}+\frac{\left(k_{p+2}\right)_{\mu} \left(k_{p+2}\right)_{\nu}}{k_{p+2}^{2}}\right)\tilde{\mathfrak{h}} & \text{verified}\\
\bottomrule
\end{array}}
\]
\caption{Classification of waves in vacuum}
\label{classificazione}
\end{table}

Consider a wave propagating along the $z$-axis.  From Eq. \eqref{sol vuoto eq1},  we have:
\begin{equation}
\label{onda asse z}
\mathfrak{h}_{\mu\nu}\left(t,z\right)=\int\frac{d^{3}\mathbf{k}}{\left(2\pi\right)^{3}}C_{\mu\nu}\left(\mathbf{k}\right)e^{i\omega_{1}\left(t-z\right)}
+\sum_{m=2}^{p+2}\int\frac{d^{3}\mathbf{k}}{\left(2\pi\right)^{3}}\left\{\frac{1}{3}\left[\frac{\eta_{\mu\nu}}{2}+\frac{\left(k_{m}\right)_{\mu}\left(k_{m}\right)_{\nu}}{k_{m}^{2}}\right]\right\}A_{m}\left(\mathbf{k}\right)e^{i\left(\omega_{m}t-k_{z}z\right)}\ ,
\end{equation}
where $k_{1}^{\mu}=\left(\omega_{1},0,0,k_{z}\right)$ and $k_{m}^{\mu}=\left(\omega_{m},0,0,k_{z}\right)$.
The oscillation mode A$_{1}$ with $k^{2}=k_{1}^{2}=0$, i.e. $\omega_{1}=q=k_{z}>0$, could have the trace of the metric perturbation not equal to zero, $\mathfrak{h}\ne0$. However, by exploiting the degrees left free, we can perform an infinitesimal transformation that makes the trace equal to zero. In fact, from Eq. \eqref{onda asse z}, considering the plane wave associated with the $\omega_{1}$ mode to $\mathbf{k}$ constant that propagates along the $z$-axis, using  the symmetry of the polarization tensor  $\epsilon_{\mu\nu}^{\text{A}_{1}}$, we have:
\begin{equation}
\mathfrak{h}_{\mu\nu}^{\text{A}_{1}}\left(t,z\right)=\epsilon_{\mu\nu}^{\text{A}_{1}}\left(\mathbf{k}\right)e^{i\omega_{1}\left(t-z\right)}=
\begin{pmatrix}
\epsilon_{00} & \epsilon_{01} & \epsilon_{02} & \epsilon_{03}\\
\epsilon_{01} & \epsilon_{11} & \epsilon_{12} & \epsilon_{13}\\
\epsilon_{02} & \epsilon_{12} & \epsilon_{22} & \epsilon_{23}\\
\epsilon_{03} & \epsilon_{13} & \epsilon_{23} & \epsilon_{33}
\end{pmatrix}
e^{i\omega_{1}\left(t-z\right)}\ ,
\end{equation}
where we have set $C_{\mu\nu}\left(\mathbf{k}\right)=\epsilon_{\mu\nu}^{\text{A}_{1}}\left(\mathbf{k}\right)$.
From the four gauge relations 
\begin{equation}
\epsilon_{\mu\nu}k^{\mu}-\frac{1}{2}\epsilon k_{\nu}=0\ ,
\end{equation}
in the $k$-space, the ten unknowns become six and precisely: $\epsilon_{01}$, $ \epsilon_{02}$, $\epsilon_{03}$, $\epsilon_{11}$, $\epsilon_{12}$ and $\epsilon$. Thus explicitly written:
\begin{equation}
\label{sist gauge}
\begin{aligned}
\begin{cases}
\epsilon_{00}+\epsilon_{30}=\frac{1}{2}\epsilon\quad &\Rightarrow \epsilon_{00}=\frac{1}{2}\epsilon-\epsilon_{30}\\
\epsilon_{01}+\epsilon_{31}=0\quad &\Rightarrow \epsilon_{31}=-\epsilon_{01}\\
\epsilon_{02}+\epsilon_{32}=0\quad&\Rightarrow \epsilon_{32}=-\epsilon_{02}\\
\epsilon_{03}+\epsilon_{33}=-\frac{1}{2}\epsilon\quad &\Rightarrow \epsilon_{33}=-\frac{1}{2}\epsilon-\epsilon_{03}\ .
\end{cases} 
\end{aligned}
\end{equation}
If we perform the infinitesimal transformation $x'^{\mu}=x^{\mu}+\xi^{\mu}$, the polarization tensor
$\epsilon_{\mu\nu}^{\textit{A}_{1}}$, at first-order in $\left| \xi \right|$, becomes \cite{Wave 4a}
\begin{equation}
\epsilon'_{\mu\nu}=\epsilon_{\mu\nu}+k_{\mu}\theta_{\nu}+k_{\nu}\theta_{\mu} \ ,
\end{equation}
if
\begin{equation}
\xi^{\mu}=i\theta^{\mu}e^{ik_{1}^{\alpha}x_{\alpha}} \ .
\end{equation}
In the mode $k_{1}^{2}=0$,  we have $\Box \xi^{\mu}=0$ that is, the gauge remains unchanged (gauge invariant) and therefore we can find a particular gauge transformation that makes  the metric perturbation, transverse and traceless. In the Fourier space, it is equivalent to impose the two conditions $ k^{\mu}\epsilon_{\mu\nu}^{\prime}=0$ and $\epsilon^{\prime}=0$, known as the TT gauge, where we have  only two degrees of freedom. So for the system \eqref{sist gauge}, we have:
\begin{equation}
\left\{
\begin{aligned}
\epsilon'_{00}&=\epsilon_{00}+2\omega_{1}\theta_{0} &\Rightarrow&\quad &\theta_{0}=-\frac{\epsilon_{00}}{2\omega_{1}}=-\frac{\epsilon}{4\omega_{1}}+\frac{\epsilon_{30}}{2\omega_{1}} \quad \text{se} \quad \epsilon'_{00}=0\\
\epsilon'_{11}&=\epsilon_{11}\\
\epsilon'_{22}&=\epsilon_{22} &\Rightarrow& &\epsilon'_{22}=-\epsilon_{11}\\
\epsilon'_{33}&=\epsilon_{33}-2\omega_{1}\theta_{3}&\Rightarrow& &\theta_{3}=-\frac{\epsilon_{33}}{2\omega_{1}}=-\frac{\epsilon}{4\omega_{1}}-\frac{\epsilon_{30}}{2\omega_{1}} \quad \text{se} \quad \epsilon'_{33}=0\\
\epsilon'_{01}&=\epsilon_{01}+\omega_{1}\theta_{1}&\Rightarrow& &\theta_{1}=-\frac{\epsilon_{01}}{\omega_{1}} \quad \text{se} \quad \epsilon'_{01}=0\\
\epsilon'_{02}&=\epsilon_{02}+\omega_{1}\theta_{2} &\Rightarrow& &\theta_{2}=-\frac{\epsilon_{02}}{\omega_{1}} \quad \text{se} \quad \epsilon'_{02}=0\\
\epsilon'_{03}&=\epsilon_{03}+\omega_{1}\theta_{3}-\omega_{1}\theta_{0} &\Rightarrow& &\epsilon'_{03}=0\\
\epsilon'_{12}&=\epsilon_{12}\\
\epsilon'_{13}&=\epsilon_{13}-k_{z}\theta_{1} &\Rightarrow& &\epsilon'_{13}=0\\
\epsilon'_{23}&=\epsilon_{23}-k_{z}\theta_{2} &\Rightarrow& &\epsilon'_{23}=0
\end{aligned}
\right.
\end{equation}
By choosing the 4-vector associated to the infinitesimal transformation as 
\begin{equation}
\theta_{\mu}=\left(-\frac{\epsilon}{4\omega_{1}}+\frac{\epsilon_{30}}{2\omega_{1}},-\frac{\epsilon_{01}}{\omega_{1}},-\frac{\epsilon_{02}}{\omega_{1}}, -\frac{\epsilon}{4\omega_{1}}-\frac{\epsilon_{30}}{2\omega_{1}}\right) \ ,
\end{equation}
the polarization tensor  $\epsilon_{\mu\nu}^{\prime\text{A}_{1}}$ becomes
\begin{equation}
\epsilon_{\mu\nu}^{\prime\text{A}_{1}}=
\begin{pmatrix}
0 & 0 & 0 & 0\\
0 & \epsilon_{11} & \epsilon_{12} & 0\\
0 & \epsilon_{12} & -\epsilon_{11} & 0\\
0 & 0 & 0 & 0
\end{pmatrix}
=\epsilon_{11}
\begin{pmatrix}
0 & 0 & 0 & 0\\
0 & 1 & 0 & 0\\
0 & 0 & -1& 0\\
0 & 0 & 0 & 0
\end{pmatrix}
+\epsilon_{12}
\begin{pmatrix}
0 & 0 & 0 & 0\\
0 & 0 & 1 & 0\\
0 & 1 & 0 & 0\\
0 & 0 & 0 & 0 
\end{pmatrix} \ ,
\end{equation}
namely for the mode A$_{1}$, we have the two transversal polarization states of  General Relativity. So the plane wave  A$_{1}$ can be written as:
\begin{equation}
\mathfrak{h}_{\mu\nu}^{\text{A}_{1}}\left(t,z\right)=\sqrt{2}\left[\epsilon_{11}\epsilon_{\mu\nu}^{\left(+\right)}+\epsilon_{12}\epsilon_{\mu\nu}^{\left(\times\right)}\right]e^{i\omega_{1}\left(t-z\right)} \ ,
\end{equation}
where we have indicated the two polarization states of  General Relativity  as
\begin{equation}
\epsilon_{\mu\nu}^{\left(+\right)}=\frac{1}{\sqrt{2}}
\begin{pmatrix}
0 & 0 & 0 & 0\\0 & 1 & 0 & 0\\0 & 0 & -1 & 0\\0 & 0 & 0 & 0
\end{pmatrix}\ ,
\qquad \epsilon_{\mu\nu}^{\left(\times\right)}=\frac{1}{\sqrt{2}}
\begin{pmatrix}
0 & 0 & 0 & 0\\0 & 0 & 1 & 0\\0 & 1 & 0 & 0\\0 & 0 & 0 & 0
\end{pmatrix} \ .
\end{equation}
Oscillation modes $A_{m}$, with $2\leq m \leq p+2$ and $k^{2}=k_{m}^{2}\ne0$, have the perturbation trace $\mathfrak{h}\neq0$, otherwise one gets A$_{m}\left(\mathbf{k}\right)=0$, that is the trivial solution $\mathfrak{h}_{\mu\nu}=0$. From Eq. \eqref{onda asse z}, we consider the plane wave related to the mode $m$ for fixed $\mathbf {k}$ that is:
\begin{equation}
\mathfrak{h}^{\text{A}_{m}}_{\mu\nu}\left(t,z\right)=\epsilon_{\mu\nu}^{\text{A}_{m}}e^{i\left(\omega_{m}t-k_{z}z\right)}=\frac{A_{m}\left(\mathbf{k}\right)}{3}
\begin{pmatrix}
\frac{1}{2}+\frac{\omega_{m}^{2}}{k_{m}^{2}} & 0 & 0 & -\frac{\omega_{m}k_{z}}{k_{m}^{2}}\\
0 & -\frac{1}{2} & 0 & 0\\ 0 & 0 & -\frac{1}{2} & 0\\ -\frac{\omega_{m}k_{z}}{k_{m}^{2}} & 0 & 0 & -\frac{1}{2}+\frac{k_{z}^{2}}{k_{m}^{2}}
\end{pmatrix}
e^{i\left(\omega_{m}t-k_{z}z\right)}\ ,
\end{equation}
where $\epsilon_{\mu\nu}^{\text{A}_{m}}\left(k\right)=\epsilon_{\mu\nu}\left(k\right)\frac{\text{A}_{m}\left(\mathbf{k}\right)}{3}$ and the harmonic gauge is verified in the Fourier space
\begin{equation}
\epsilon_{\mu\nu}k^{\mu}-\frac{1}{2}\epsilon k_{\nu}=0 \ .
\end{equation}
The harmonic gauge, in general, is not an invariant gauge, and for a general gauge transformation $x^{\prime\mu}=x^{\mu}+\xi^{\mu}$, it  turns out to be:
\begin{equation}
\partial_{\mu}\left(h^{\prime\mu\nu}-\frac{1}{2}\eta^{\mu\nu}h^{\prime}\right)=\partial_{\mu}\left(h^{\mu\nu}-\frac{1}{2}\eta^{\mu\nu}h\right)-\Box\xi^{\nu} \ .
\end{equation}
Our transformation maps harmonic gauge into harmonic gauge if and only if $\xi^{\nu}$ is the solution of equation $\Box\xi^{\nu}=0$. So we will have as a solution 
\begin{equation}
\xi^{\nu}\left(x\right)=\int\frac{d^{3}\mathbf{k}}{\left(2\pi\right)^{3}}\theta^{\mu}\left(\mathbf{k}\right)e^{ik_{m}^{\alpha}x_{\alpha}}\Leftrightarrow k_{m}^{2}=0 \ .
\end{equation}
But, for massive modes $k_{m}^{2}\neq0$, we have: 
\begin{equation}
\Box\xi^{\nu}\left(x\right)=\int\frac{d^{3}\mathbf{k}}{\left(2\pi\right)^{3}}\theta^{\mu}\left(\mathbf{k}\right)\left(-k_{m}^{2}\right)e^{ik_{m}^{\alpha}x_{\alpha}}\neq0 \ ,
\end{equation}
unless  $\theta^{\nu}=0$, namely the identical transformation. For the massive modes there is no gauge transformation that leaves the harmonic gauge unchanged, that is  the polarization tensor $\epsilon_{\mu\nu}^{\text{A}_{m}}$ cannot be modified. Hence it is not possible to make the polarization tensor for these massive modes neither completely spatial nor traceless. Expressing it as a function of a suitable orthonormal polarization basis,  we have:
\begin{multline}
\epsilon_{\mu\nu}^{\text{A}_{m}}=\frac{A_{m}\left(\mathbf{k}\right)}{3}
\begin{pmatrix}
\frac{1}{2}+\frac{\omega_{m}^{2}}{k_{m}^{2}} & 0 & 0 & -\frac{\omega_{m}k_{z}}{k_{m}^{2}}\\
0 & -\frac{1}{2} & 0 & 0\\ 0 & 0 & -\frac{1}{2} & 0\\ -\frac{\omega_{m}k_{z}}{k_{m}^{2}} & 0 & 0 & -\frac{1}{2}+\frac{k_{z}^{2}}{k_{m}^{2}}
\end{pmatrix}
=\frac{A_{m}\left(\mathbf{k}\right)}{3}\Biggl\{
\left(\frac{1}{2}+\frac{\omega_{m}^{2}}{k_{m}^{2}}\right)
\begin{pmatrix}
1 & 0 & 0 & 0\\0 & 0 & 0 & 0\\0 & 0 & 0 & 0\\0 & 0 & 0 & 0
\end{pmatrix}\\
+\left(-\frac{1}{2}\right)
\begin{pmatrix}
0 & 0 & 0 & 0\\0 & 1 & 0 & 0\\0 & 0 & 1 & 0\\0 & 0 & 0 & 0
\end{pmatrix}
+\left(-\frac{\omega_{m}k_{z}}{k_{m}^{2}}\right)
\begin{pmatrix}
0 & 0 & 0 & 1\\0 & 0 & 0 & 0\\0 & 0 & 0 & 0\\1 & 0 & 0 & 0
\end{pmatrix}
+\left(-\frac{1}{2} +\frac{k_{z}^{2}}{k_{m}^{2}}\right)
\begin{pmatrix}
0 & 0 & 0 & 0\\0 & 0 & 0 & 0\\0 & 0 & 0 & 0\\0 & 0 & 0 & 1
\end{pmatrix}\Biggr\} \ .
\end{multline}
Indicating the four polarization states as
\begin{equation}
\begin{matrix}
&\epsilon_{\mu\nu}^{\left(\text{TT}\right)}=\begin{pmatrix}
1 & 0 & 0 & 0\\0 & 0 & 0 & 0\\0 & 0 & 0 & 0\\0 & 0 & 0 & 0
\end{pmatrix}\ ,
& \epsilon_{\mu\nu}^{\left(\text{TS}\right)}=\frac{1}{\sqrt{2}}
\begin{pmatrix}
0 & 0 & 0 & 1\\0 & 0 & 0 & 0\\0 & 0 & 0 & 0\\1 & 0 & 0 & 0
\end{pmatrix}\ ,\\ \\
&\epsilon_{\mu\nu}^{\left(1\right)}=\frac{1}{\sqrt{2}}
\begin{pmatrix}
0 & 0 & 0 & 0\\0 & 1 & 0 & 0\\0 & 0 & 1 & 0\\0 & 0 & 0 & 0
\end{pmatrix}\ ,
& \epsilon_{\mu\nu}^{\left(L\right)}=
\begin{pmatrix}
0 & 0 & 0 & 0\\0 & 0 & 0 & 0\\0 & 0 & 0 & 0\\0 & 0 & 0 & 1
\end{pmatrix} \ ,
\end{matrix}
\end{equation}
we can express  A$_{m}$ plane waves as:
\begin{equation}
\begin{split}
\mathfrak{h}_{\mu\nu}^{\text{A}_{m}}\left(t,z\right)=&\Biggl[\frac{1}{3}\left(\frac{1}{2}+\frac{\omega_{m}^{2}}{k_{m}^{2}}\right)\epsilon_{\mu\nu}^{\left(\text{TT}\right)}\\&-\frac{\sqrt{2}\omega_{m}k_{z}}{3k_{m}^{2}}\epsilon_{\mu\nu}^{\left(\text{TS}\right)}-\frac{\sqrt{2}}{6}\epsilon_{\mu\nu}^{\left(1\right)}+\frac{1}{3}\left(-\frac{1}{2} +\frac{k_{z}^{2}}{k_{m}^{2}}\right)\epsilon_{\mu\nu}^{\left(L\right)}\Biggr]\text{A}_{m}\left(\mathbf{k}\right)
e^{i\left(\omega_{m}t-k_{z}z\right)} \ .
\end{split} 
\end{equation}
So the general solution for a wave that propagates along the $z$-axis considering the $(p+2)$ oscillation modes $\omega_{m}$  becomes:
\begin{equation}
\begin{split}
\mathfrak{h}_{\mu\nu}\left(t,z\right)=&\int\frac{d^{3}\mathbf{k}}{\left(2\pi\right)^{3}}\left[\sqrt{2}\epsilon_{11}\epsilon_{\mu\nu}^{\left(+\right)}+\sqrt{2}\epsilon_{12}\epsilon_{\mu\nu}^{\left(\times\right)}\right]e^{i\omega_{1}\left(t-z\right)}\\&+\sum_{m=2}^{p+2}\int\frac{d^{3}\mathbf{k}}{\left(2\pi\right)^{3}}\Biggl[\frac{1}{3}\left(\frac{1}{2}+\frac{\omega_{m}^{2}}{k_{m}^{2}}\right)\epsilon_{\mu\nu}^{\left(\text{TT}\right)}-\frac{\sqrt{2}\omega_{m}k_{z}}{3k_{m}^{2}}\epsilon_{\mu\nu}^{\left(\text{TS}\right)}\\&-\frac{\sqrt{2}}{6}\epsilon_{\mu\nu}^{\left(1\right)}+\frac{1}{3}\left(-\frac{1}{2} +\frac{k_{z}^{2}}{k_{m}^{2}}\right)\epsilon_{\mu\nu}^{\left(L\right)}\Biggr]\text{A}_{m}\left(\mathbf{k}\right)e^{i\left(\omega_{m}t-k_{z}z\right)} \ .
\end{split} 
\end{equation}
expressed with respect to the polarization basis $\epsilon_{\mu\nu}^{\left(+\right)}$,  $\epsilon_{\mu\nu}^{\left(\times\right)}$, $\epsilon_{\mu\nu}^{\left(\text{TT}\right)}$, $\epsilon_{\mu\nu}^{\left(\text{TS}\right)}$, $\epsilon_{\mu\nu}^{\left(1\right)}$, $\epsilon_{\mu\nu}^{\left(\text{L}\right)}$.
In terms of amplitudes,  the solution can be written as:
\begin{equation}
\begin{split}
{h}_{\mu\nu}\left(t,z\right)= &\text{A}^{\left(+\right)}\left(t-z\right)\epsilon_{\mu\nu}^{\left(+\right)}+\text{A}^{\left(\times\right)}\left(t-z\right)\epsilon_{\mu\nu}^{\left(\times\right)}\\
&+\sum_{m=2}^{p+2}\biggl[\text{A}_{m}^{\left(TT\right)}\left(t-v_{G_{m}}z\right)\epsilon_{\mu\nu}^{\left(TT\right)}+\text{A}_{m}^{\left(TS\right)}\left(t-v_{G_{m}}z\right)\epsilon_{\mu\nu}^{\left(TS\right)}\\
&+\text{A}_{m}^{\left(1\right)}\left(t-v_{G_{m}}z\right)\epsilon_{\mu\nu}^{\left(1\right)}+\text{A}_{m}^{\left(L\right)}\left(t-v_{G_{m}}z\right)\epsilon_{\mu\nu}^{\left(L\right)}\biggr] \ ,
\end{split}
\end{equation}
where $v_{G_{m}}$ is the group velocity, related to the massive mode $m$, defined below in Eq. (\ref{velocita gruppo}).
The polarization tensors have been chosen  to form an orthonormal basis, that is  they have to  verify the following relation:
\begin{equation}
\text{Tr}\left\{\epsilon^{\left(i\right)}\epsilon^{*\left(j\right)}\right\}\equiv\epsilon_{\mu\nu}^{\left(i\right)}\epsilon^{*\left(j\right)\mu\nu}=\delta^{ij} \quad \text{with} \quad i,j\in\left\{+,\times,\text{TT},\text{TS}, 1,\text{L}\right\}
\end{equation}
The three polarization states $\epsilon_{\mu\nu}^{\left(+\right)}$,  $\epsilon_{\mu\nu}^{\left(\times\right)}$ ed $\epsilon_{\mu\nu}^{\left(1\right)}$ are transversal and verify the relation $\epsilon_{\mu\nu}k^{\mu}=0$:
\begin{equation}
\epsilon_{\mu\nu}^{\left(+\right)}k^{\mu}=0\ , \qquad \epsilon_{\mu\nu}^{\left(\times\right)}k^{\mu}=0\ , \qquad \epsilon_{\mu\nu}^{\left(1\right)}k^{\mu}=0 \ ,
\end{equation}
while the remaining three polarization states $\epsilon_{\mu\nu}^{\left(\text{TT}\right)}$, $\epsilon_{\mu\nu}^{\left(\text{TS}\right)}$, $\epsilon_{\mu\nu}^{\left(\text{L}\right)}$ are not transversal:
\begin{equation}
\epsilon_{\mu\nu}^{\left(\text{TT}\right)}k^{\mu}\neq0\ , \qquad \epsilon_{\mu\nu}^{\left(\text{TS}\right)}k^{\mu}\neq0\ , \qquad \epsilon_{\mu\nu}^{\left(\text{L}\right)}k^{\mu}\neq0 \ .
\end{equation}
In summary,  there are $6$ polarization states.
In order to study the helicity of such waves, we see how the polarization basis  $\{\epsilon_{\mu\nu}^{\left(+\right)}$, $ \epsilon_{\mu\nu}^{\left(\times\right)}, \epsilon_{\mu\nu}^{\left(\text{TT}\right)}$, $\epsilon_{\mu\nu}^{\left(\text{TS}\right)}, \epsilon_{\mu\nu}^{\left(1\right)}, \epsilon_{\mu\nu}^{\left(\text{L}\right)}\}$ under a rotation of an angle $\varphi$ around the $z$-axis:
\begin{equation}
R_{\mu}^{\ \nu}=
\begin{pmatrix}
1 & 0 & 0 & 0\\
0 & \cos\varphi & \sin\varphi & 0\\
0 & -\sin\varphi & \cos\varphi & 0\\
0 & 0 & 0 & 1
\end{pmatrix} \ ,
\end{equation}
that is $\tilde{\epsilon}_{\mu\nu}=R_{\mu}^{\ \rho}R_{\nu}^{\ \sigma}\epsilon_{\rho\sigma}$.
The four polarizations $\epsilon_{\mu\nu}^{\left(\text{TT}\right)}$, $\epsilon_{\mu\nu}^{\left(\text{TS}\right)}$, $\epsilon_{\mu\nu}^{\left(1\right)}$ and $\epsilon_{\mu\nu}^{\left(\text{L}\right)}$ remain unchanged under rotations and therefore the waves  A$_{m}$ with $2\leq m \leq p+2$ have helicity equal to zero.\\
If we introduce two further polarization states, called circular,
\begin{equation}
\epsilon_{\mu\nu}^{\left(\text{R}\right)}=\frac{1}{\sqrt{2}}\left(\epsilon_{\mu\nu}^{\left(+\right)}+i\epsilon_{\mu\nu}^{\left(\times\right)}\right) \ ,
\end{equation}
and
\begin{equation}
\epsilon_{\mu\nu}^{\left(\text{L}\right)}=\frac{1}{\sqrt{2}}\left(\epsilon_{\mu\nu}^{\left(+\right)}-i\epsilon_{\mu\nu}^{\left(\times\right)}\right) \ ,
\end{equation}
we see that, under  rotation, they  transform as:
\begin{equation}
\epsilon_{\mu\nu}^{\prime{\text{R}\choose\text{L}}}=e^{\pm 2i\varphi}\epsilon_{\mu\nu}^{\text{R}\choose\text{L}} \ ,
\end{equation}
that is, waves like A$ _ {1}$ have two-helicity because they are the standard ones of General Relativity.

It is possible to prove that our Lagrangian \eqref{laglag} is conformally equivalent to Einstein's theory with $(p+1)$ appropriate scalar fields \cite{Var derivative}. Setting $k_{m}^{2}=M_{m}^{2}$ for the  waves  A$_{m}$,  
the dispersion relation becomes $\omega_{m}\left(q\right)=\sqrt{M_{m}^{2}+q^{2}}$. We can interpret the $(p+1)$ oscillation modes as massive scalar fields of mass $M$  with four polarization states, one transverse $\epsilon_{\mu\nu}^{\left(1\right)}$ and three longitudinal $\epsilon_{\mu\nu}^{\left(\text{TT}\right)}$, $\epsilon_{\mu\nu}^{\left(\text{TS}\right)}$, $\epsilon_{\mu\nu}^{\left(L\right)}$ with helicity equal to zero.
 We can associate a massless tensor field to the wave A$_{1}$, with two transverse polarization states with helicity two. The wave associated with massless mode A$_{1}$ has velocity $c$, whereas  A$_{m}$ waves have a velocity other than $c$,  due to the dispersion law. That is, if we consider the wave packet  associated with such modes, the group velocity is:
\begin{equation}\label{velocita gruppo}
v_{{G}_{m}}=\frac{d\omega_{m}\left(q\right)}{dq}=\frac{\sqrt{\omega_{m}^{2}-M_{m}^{2}}}{\omega_{m}} \ ,
\end{equation}
which allows us to associate a velocity to the wave A$_{m}$ and therefore to  the scalar field.
A summary of  polarizations and  helicity states is reported  in  Table \ref{pol ed eli}.
\begin{table}[H]
\begin{equation*}
\begin{array}{ccccc}
\toprule
\text{Mode}  & \text{Dispersion relation} & \text{Polarization} & \text{Helicity} & \text{Mass of the associated field}\\
\midrule
\text{A}_{1}  & \omega_{1}=q & \epsilon_{\mu\nu}^{\left(+\right)},\epsilon_{\mu\nu}^{\left(\times\right)} & 2 & 0 \\
\midrule
\text{A}_{2} & \omega_{2}=\sqrt{k_{2}^{2}+q^{2}} & \epsilon_{\mu\nu}^{\left(\text{TT}\right)}, \epsilon_{\mu\nu}^{\left(\text{TS}\right)}\epsilon_{\mu\nu}^{\left(1\right)},\epsilon_{\mu\nu}^{\left(L\right)} & 0 & M_{2}=k_{2}\\
\midrule
\vdots & \vdots & \vdots & \vdots & \vdots\\
\midrule
\text{A}_{p+2} & \omega_{p+2}=\sqrt{k_{p+2}^{2}+q^{2}} & \epsilon_{\mu\nu}^{\left(\text{TT}\right)}, \epsilon_{\mu\nu}^{\left(\text{TS}\right)}\epsilon_{\mu\nu}^{\left(1\right)},\epsilon_{\mu\nu}^{\left(L\right)} & 0 & M_{p+2}=k_{p+2}\\
\bottomrule
\end{array}
\end{equation*}
\caption{Polarizations and helicity states}
\label{pol ed eli}
\end{table}

\section{Conclusions}
Properties of gravitational waves   provide fundamental  information for any theory of  gravity.
In particular, they allow to set  constraints on the gravitational Lagrangian considering the multipolar radiation,  polarization and helicity states  \cite{DLSC, DLDM} also considering, indirectly, astrophysical dynamics \cite{DeLa}. Here we have taken into account a generic higher-order gravitational Lagrangian density $L_{g}=(R+a_{0}R^{2}+\sum_{k=1}^{p} a_{k}R\Box^{k}R)\sqrt{-g}$. We   perturbed  the metric $g_{\mu\nu}$ with respect to the  flat  Minkowski spacetime $\eta_{\mu\nu}$ and thus, we  obtained the linearized equations in the perturbed metric $h_{\mu\nu}$. 
The solutions are gravitational waves with $(p+2)$ normal modes of helicity 0 and 2  with 6 polarization states, three transverse and three longitudinal. Here $p$ is the order of the theory so the result is completely general for theories of any order. It is important to stress the fact that, in four dimensions, the maximal allowed number of polarization state is always 6 for any theory of gravity. This fact has a deep intrinsic meaning that can  be related  to the fundamental structure of spacetime and the degrees of freedom of gravitational field (see also \cite{BCDN,abedi}).

In principle, 
the  emitted power from  a gravitational radiating source can be related to the   gravitational  
energy-momentum pseudo-tensor (see \cite{CCT1,CCLS, CCT2} for details in metric and teleparallel gravity). Then features of sources and further gravitational modes could be strictly related. In this sense, the so called {\it multimessenger astrophysics} is  a powerful tool to discriminate among concurring gravitational theories  (see \cite{horn1,horn2,horn3}).

In particular, relating  terrestrial laser interferometers like LIGO (Livingston and Hanford, USA), VIRGO (Cascina, Italy), GEO 600 (Hannover, Germany),  Tama 300 (Mitaka, Japan),  KAGRA (Japan), LIGO-India (India) and,  in principle,  the space interferometer LISA, could be  the best approach to detect or exclude   possible further gravitational modes.  

\section*{Acknowledgements}
 SC is supported in part by the INFN sezione di Napoli, {\it iniziative specifiche} TEONGRAV and QGSKY.
The  article is also based upon work from COST action CA15117 (CANTATA),
supported by COST (European Cooperation in Science and Technology).


\begin{thebibliography}{99}

\bibitem{luongo} O. Luongo, M. Muccino, \emph{Speeding up the universe using dust with pressure},  
Phys. Rev. D \textbf{98}, 103520 (2018).

\bibitem{agostino} S. Capozziello, Rocco D'Agostino, O. Luongo, 	\emph{Cosmic acceleration from a single fluid description}, Phys. Dark  Univ. \textbf{20}, 1-12, (2018).

\bibitem{dunsby} P. K. S. Dunsby, O. Luongo, L. Reverberi, 
\emph{Dark Energy and Dark Matter from an additional adiabatic fluid}, Phys. Rev. D \textbf{94}, 083525 (2016).
 
\bibitem{aviles} A. Aviles, L. Bonanno, O. Luongo, H. Quevedo, 
\emph{Holographic dark matter and dark energy with second order invariants}, 
Phys. Rev. D \textbf{84}, 103520, (2011).

\bibitem{quevedo} O. Luongo, H. Quevedo,	
\emph{A Unified Dark Energy Model from a Vanishing Speed of Sound with Emergent Cosmological Constant}, Int. J. Mod. Phys. D \textbf{23}, 1450012, (2014).

\bibitem{bravetti} A. Bravetti, O. Luongo, 	
\emph{Dark energy from geometrothermodynamics},
Int Jour Geom Meth Mod Phys, \textbf{11},  1450071 (2014).

\bibitem{mancini} S. Capozziello, O. Luongo, S. Mancini, 	
\emph{Cosmological dark energy effects from entanglement}, Phys. Lett. A \textbf{377}, 1061  (2013).  

\bibitem{luongo2} S Capozziello, O Luongo, Int.J.Mod.Phys. D \textbf{27}  1850029, (2017).

\bibitem{luongo3} S Capozziello, O Luongo, 	
\emph{Dark energy from entanglement entropy},  Int. J. Theor. Phys.  	
 \textbf{52}, 2698  (2013).

\bibitem{luongo4} S Capozziello, O Luongo, \emph{Entanglement inside the cosmological apparent horizon} Phys. Lett. A \textbf{378} 2058  (2014).

\bibitem{luongo5} S. Capozziello, O. Luongo, \emph{Entangled states in quantum cosmology and the interpretation of Lambda, Entropy} \textbf{13}, 528, (2011)  

\bibitem{dunsby2} P. K. S. Dunsby, O. Luongo,  \emph{On the theory and applications of modern cosmography},
  Int. J. Geom. Meth. Mod. Phys. \textbf{13},  1630002 (2016).


\bibitem{Ext Theory} S. Capozziello, M. De Laurentis, \emph{Extended Theories of Gravity}, Phys. Rept. \textbf{ 509}, 167  (2011).


\bibitem{CFA} S. Capozziello, V. Faraoni, \emph{Beyond Einstein Gravity: A Survey of Gravitational Theories for Cosmology and Astrophysics}, Springer, Fundam. Theor. Phys. 170, Dordrecht  (2010).

\bibitem{sergey}
  S.~Nojiri and S.~D.~Odintsov,
  \emph{Unified cosmic history in modified gravity: from F(R) theory to Lorentz non-invariant models}
  Phys.\ Rept.\  {\bf 505}, 59  (2011). 
  
\bibitem{vasilis}
  S.~Nojiri, S.~D.~Odintsov and V.~K.~Oikonomou,
  \emph{Modified Gravity Theories on a Nutshell: Inflation, Bounce and Late-time Evolution}
  Phys.\ Rept.\  {\bf 692}, 1 (2017).
  
  \bibitem{felix}
  M.~De Laurentis and A.~J.~Lopez-Revelles,
  \emph{Newtonian, Post Newtonian and Parameterized Post Newtonian limits of f(R, G) gravity}
  Int.\ J.\ Geom.\ Meth.\ Mod.\ Phys.\  {\bf 11},  1450082 (2014).
  
 \bibitem{bini}
  C.~Cherubini, D.~Bini, S.~Capozziello and R.~Ruffini,
  \emph{Second order scalar invariants of the Riemann tensor: Applications to black hole space-times}
  Int.\ J.\ Mod.\ Phys.\ D {\bf 11}, 827  (2002).
  
\bibitem{birrell}
N.D. Birrel and P.C.W. Davies {\it Quantum Fields in Curved Space},  Cambridge University Press, Cambridge (1982).

\bibitem{vilkovisky}
G.A. Vilkovisky,  {\it Effective action in quantum gravity},  Class. Quant. Grav. {\bf 9}, 895 (1992).

\bibitem{breno1}
B. L. Giacchini, I. L. Shapiro,
\emph{Light bending in $F[g(\Box)R]$ extended gravity theories}, Phys. Lett. B \textbf{780}, 54 (2018).

\bibitem{breno2}
B. L. Giacchini, T. de Paula Netto, 
\emph{Weak-field limit and regular solutions in polynomial higher-derivative gravities}, 
e-Print: arXiv:1806.05664 [gr-qc]  (2018).

\bibitem{breno3}
B. L. Giacchini, T. de Paula Netto, 
\emph{Effective delta sources and regularity in higher-derivative and ghost-free gravity},
e-Print: arXiv:1809.05907 [gr-qc]  (2018).

\bibitem{annalen}
  S.~Capozziello and M.~De Laurentis,
 {\it The dark matter problem from f(R) gravity viewpoint,}
  Annalen Phys.\  545 {\bf 524} (2012).
  
   
\bibitem{HT} R. A. Hulse and J. H. Taylor, \emph{Discovery of a pulsar in a binary system}, Astrophys. J. \textbf{195}, L51 (1975).

\bibitem{TW} J. H. Taylor and J. M. Weisberg, \emph{A new test of general relativity - Gravitational radiation and the binary pulsar PSR 1913+16}, Astrophys. J. \textbf{253}, 908 (1982).

\bibitem{ABGW}B. P. Abbott et al., \emph{Observation of Gravitational Waves from a Binary Black Hole Merger}, Phys. Rev. Lett. \textbf{116}, 061102 (2016).

\bibitem{WLR} C. M. Will,  \emph{The Confrontation between General Relativity and Experiment}, Living Rev. Relativ. \textbf{9}, 3 (2006).

\bibitem{BCDN} C. Bogdanos, S. Capozziello, M. De Laurentis and S. Nesseris, \emph{Massive, massless
and ghost modes of gravitational waves from higher-order gravity}, Astropart. Phys. \textbf{34}, 236 (2010).

\bibitem{CCT1}S. Capozziello, M. Capriolo and M. Transirico, \emph{The gravitational energy-momentum
pseudo-tensor of higher-order theories of gravity}, Ann. Phys. \textbf{525}, 1600376 (2017).

\bibitem{Sixt order} S. Gottlober, H.J. Schmidt and A. A. Starobinsky, \emph{Sixth Order Gravity and Conformal Transformations}, Class. Quant. Grav. \textbf{7}, 893 (1990).

\bibitem{Var derivative}  H.J. Schmidt, \emph{ Variational derivatives of arbitrarily high order and multi-inflation cosmological models}, Class. Quantum Grav. \textbf{7}, 1023 (1990).

\bibitem{Weinb}S. Weinberg, \emph{Gravitation and Cosmology}, Wiley, New York (1972).

\bibitem{CARLN} S. M. Carroll, \emph{Lecture Notes on General Relativity} \url{arXiv:gr-qc/9712019} (1997).

\bibitem{CST}S. Capozziello, A. Stabile and A. Troisi, \emph{The Post-Minkowskian Limit of f(R)-gravity}, Int. J. Theor. Phys. \textbf{49}, 1251, (2010).

\bibitem{Limit New} I. Quandt, H.J. Schmidt, \emph{The Newtonian limit of fourth and higher order gravity},
Astron. Nach. \textbf{312},  97,  (1991). 

\bibitem{NLFO}H.J. Schmidt, \emph{The Newtonian limit of fourth-order gravity}, Astron. Nachr. \textbf{307}, 339 (1986).

\bibitem{HSJ} H.J. Schmidt, \emph{Fourth order gravity: equations, history, and applications to cosmology}, Int.
J. Geom. Methods Mod. Phys. \textbf{4}, 209 (2007).

\bibitem{LL}L.D. Landau and E.M. Lifshitz, \emph{The Classical Theory of Fields}, Pergamon Press, Oxford (1971).

\bibitem{SC} B. Schutz, \emph{A First Course in General Relativity}, Cambridge University Press, New York (2009).

\bibitem{RINDLER}W. Rindler, \emph{Relativity:
Special, General, and Cosmological}, Oxford University Press, New York (2006).

\bibitem{MTW}C. W. Misner, K.S. Thorne, J.A. Wheeler, \emph{Gravitation}, Freeman and Co., New York
(1971).

\bibitem{Wave 4a} S. Capozziello, A. Stabile, \emph{Gravitational waves in fourth order gravity}, Astrophys. Space Sci \textbf{358}, 27 (2015).

\bibitem{STELLE}K. S. Stelle, \emph{Classical Gravity with Higher Derivatives}, Gen. Rel. Grav. \textbf{9}, 353 (1978).



\bibitem{DLSC}M. De Laurentis and S. Capozziello, \emph{Quadrupolar gravitational radiation as a testbed
for f(R)-gravity}, Astropart. Phys. \textbf{35}, 257 (2011).

\bibitem{DLDM} M. De Laurentis, I. De Martino, \emph{Testing f(R)-theories using the first time derivative
of the orbital period of the binary pulsars}, Mon. Not. Roy. Astron. Soc. \textbf{431}, 741 (2014).

\bibitem{DeLa}
  M.~De Laurentis, Z.~Younsi, O.~Porth, Y.~Mizuno and L.~Rezzolla,
  {\it Test-particle dynamics in general spherically symmetric black hole spacetimes,}
  Phys.\ Rev.\ D {\bf 97} 104024 (2018).
  
\bibitem{abedi}
  H.~Abedi and S.~Capozziello,
  {\it Gravitational waves in modified teleparallel theories of gravity,}
  Eur.\ Phys.\ J.\ C {\bf 78}, 474  (2018).

\bibitem{CCLS}Y. Cai, S. Capozziello, M. De Laurentis and E. N. Saridakis, \emph{f(T) teleparallel gravity
and cosmology}, Rept. Prog. Phys. \textbf{79}, 106901 (2016).

\bibitem{CCT2}S. Capozziello, M. Capriolo and M. Transirico, \emph{The Gravitational Energy-Momentum
Pseudotensor: the cases f(R) and f(T) Gravity}, Int. J. Geom. Methods Mod. Phys. \textbf{15}, 1850164 (2018).
  
  
\bibitem{horn1}
  Y.~Gong, S.~Hou, E.~Papantonopoulos and D.~Tzortzis,
 {\it Gravitational waves and the polarizations in Horava gravity after GW170817}
  Phys.\ Rev.\ D {\bf 98},   104017 (2018).
  
\bibitem{horn2}
  Y.~Gong, S.~Hou, D.~Liang and E.~Papantonopoulos,
  {\it Gravitational waves in Einstein-aether and generalized TeVeS theory after GW170817}
  Phys.\ Rev.\ D {\bf 97},  084040 (2018).
 
\bibitem{horn3}
  Y.~Gong, E.~Papantonopoulos and Z.~Yi,
 {\it Constraints on scalar-tensor theory of gravity by the recent observational results on gravitational waves,}
  Eur.\ Phys.\ J.\ C {\bf 78},   738 (2018).
  
  


\end{thebibliography}
\end{document}